\documentclass[prl, twocolumn]{revtex4}
\usepackage{graphicx}
\usepackage{amssymb}
\usepackage[latin1]{inputenc}

\begin{document}

\title{Band topology and quantum spin  Hall effect in bilayer graphene}

\author{E. Prada$^1$, P. San-Jose$^2$ and  L. Brey$^1$ }
\affiliation{ $^1$Instituto de Ciencia de Materiales de Madrid,
(CSIC), Cantoblanco, 28049 Madrid, Spain\\
$^2$Instituto de Estructura de la Materia (CSIC), Serrano 123, 28006
Madrid, Spain} \author{H.A. Fertig} \affiliation{ Department of
Physics, Indiana University, Bloomington IN 47405}

\date{\today}

\keywords{Graphene nanoribbons \sep Electronic properties \sep Transport properties \sep Heterostructures}
\pacs{61.46.-w, 73.22.-f, 73.63.-b}

\begin{abstract}
We consider bilayer graphene in the presence of spin orbit coupling, to assess its behavior as a topological insulator. The first Chern number $n$ for the energy bands of single and bilayer graphene is computed and compared. It is shown that for a given valley and spin, $n$ in a bilayer is doubled with respect to the monolayer. This implies that bilayer graphene will have twice as many edge states as single layer graphene, which we confirm with numerical calculations and analytically in the case of an armchair terminated surface. Bilayer graphene is a weak topological insulator, whose surface spectrum is susceptible to gap opening under spin-mixing perturbations. We also assess the stability of the associated topological bulk state of bilayer graphene under various perturbations.  Finally, we consider an intermediate situation in which only one of the two layers has spin orbit coupling, and find that although individual valleys have non-trivial Chern numbers, the spectrum as a whole is not gapped, so that the system is
not a topological insulator.
\end{abstract}

\maketitle

\section{\label{sec:intro} Introduction}
The study  and characterization of the
properties of topological insulators has sparked considerable interest \cite{Hasan_2010,Qi_2010}.  A
topological insulator has a bulk energy gap separating the occupied electronic bands from the empty ones.
However, the conduction and valence
bands in the bulk are inverted with respect to their energy position in
the vacuum. This necessarily results in the existence of localized
surface states that cross the energy gap and are protected by time
reversal symmetry. The electronic and magnetic properties of the
surface states of a three-dimensional topological insulator are chiral, and are governed by a two
dimensional Dirac Hamiltonian.

In a two dimensional topological insulator the chiral surface states
become helical one dimensional edge states, and at a given edge,
states with opposite spin orientations transport charge in opposite
spatial directions\cite{note1}.
In the absence of time reversal symmetry breaking, the helicity of
the edge states prevents backscattering between states in the same
edge and the Hall conductivity per spin becomes quantized, although
with different sign for opposite spins. Note that this system does
not violate time reversal symmetry because, when reversing the time
direction, both the spin and the Hall conductivity per spin reverse.
In this way the total Hall conductivity of the system vanishes. As
pointed out by Thouless {\it et al.} \cite{Thouless_1982}, the
quantization of the Hall conductivity is a consequence of the
topology of the band structure. Thouless {\it et al.} showed that
the Hall conductivity should be an integer in units of $e^2 / h$,
i.e., a topological invariant called the first Chern number. Because
the Hall conductivity per spin is quantized, two-dimensional
topological insulators are also called quantum spin Hall systems.
The quantum spin Hall effect was theoretically predicted
\cite{Bernevig_2006} to occur, and later experimentally observed
\cite{Konig_2007}, in HgTe quantum wells confined by CdTe barriers.

Graphene, a two-dimensional carbon crystal \cite{Castro_Neto_RMP}, has also been proposed to be a two-dimensional quantum
spin Hall system \cite{Haldane_1988,Kane_2005a,Kane_2005b} when the intrinsic spin orbit interaction is taken into account. Spin
orbit coupling (SOC) in graphene opens a gap at the Dirac points and the
system becomes a topological insulator. However, due to the $\pi$ character of the graphene bands close to the Fermi energy, the
opened gap is proportional to the square of the intra-atomic spin orbit
coupling constant divided by the  energy difference between the $s$
and $p$ orbitals of graphene \cite{Huertas_2006,Min_2006,Yao_2007}.
Since carbon is a very light atom, the spin orbit gap
is very small, $\sim 10 ^{-3}$meV. Consequently, the particular
transport properties of graphene as a topological insulator may only
be observed in extremely clean samples and at extremely low
temperatures. Graphene, nevertheless, is a pedagogical toy model for
analyzing properties of two-dimensional topological
insulators \cite{Hasan_2010}.

In a Bernal stacked graphene bilayer, the conduction and the
valence bands touch at two inequivalent points of the Brillouin zone and the
system is a semimetal \cite{McCann_2006}. SOC in bilayer graphene also opens a gap in the band structure \cite{Gelderen_2010}.

The spin-orbit interaction in bilayer graphene is larger, by about
one order of magnitude, than in a single layer of graphene, due to
the mixing of $\pi$ and $\sigma$ bands by the interlayer hopping
\cite{Guinea_SO_2010,Liu_2010}. Nevertheless, the spin-orbit induced
gap in the bilayer is small and difficult to detect experimentally.
As mentioned above, our main interest in this work is the use of
graphene based structures to understand properties of topological
insulators. In particular, we use bilayer graphene as a model to
analyze the coupling between two topological insulators.

Our aim in this work is to study the topological properties of a
graphene bilayer in the presence of SOC. In graphene and bilayer
graphene, the low energy properties can be described by 2$\times$2
Hamiltonians and for each momentum $\vec{p}$, the wavefunctions have
a spinor form. In undoped systems, the ground state of the system
can be characterized by an unit vector field, $\vec{h}(\vec{p})$,
that indicates, at each point of the reciprocal space, the
expectation value of the orientation of the pseudospin. In the
reciprocal space $\vec{h}$ has the form of a topological object. We
find that in the case of graphene $\vec{h}$ has the form of a meron,
and in the case of bilayer graphene it takes the form of a double
vortex meron. Thus, we find that the first Chern number of bilayer
graphene is twice that of the monolayer one, and consequently the
number of edge states is also doubled. In bilayers, edge states are
not Kramer-protected against backscattering, which give the
topological insulating phase a weak character. We also analyze the
stability of the insulating bulk topology in bilayer graphene with
respect to bias  voltage, staggered sublattice potential and
trigonal warping effects. Finally, we study a bilayer graphene
system in which only a single layer has SOC. We find that this
system has a finite Chern number, but is a zero gap semiconductor
for which  no surface states are possible.

\section{Graphene}

Carbon atoms in graphene  are covalently bonded and arranged in a
honeycomb lattice, which is composed of two triangular sublattices
$A$ and $B$. The low energy properties in graphene are mainly
determined by the $\pi$ orbitals. A  tight-binding Hamiltonian with
hopping $\gamma_0$ between nearest-neighbors appropriately describes
its band structure.  In graphene the intrinsic SOC does not break
the inversion symmetry of the honeycomb lattice and the electronic
bands are spin degenerate. In addition, since the intrinsic SOC in
graphene is a second order effect, the $z$-component of the electron
spin commutes with the Hamiltonian, and the bands can be indexed by
$s_z=\pm 1$, the up/down electron spin component perpendicular to
the graphene layer. The corresponding tight binding Hamiltonian of
graphene is
\begin{equation}
H = - \gamma _0  \!  \! \! \! \sum _{<i,j>,s_z}
a_{i,s_z} ^+ b
_{j,s_z} + i t _{so} \! \! \! \! \sum _{<< i,j>>,s_z} s_z \nu _{ij} c ^+
_{i,s_z} c_{j,s_z} .
\end{equation}
Here $a_{i,s_z}(b_{i,s_z})$ annihilates an electron on sublattice
$A(B)$ at site $i$ and  spin $s_z$, $t_{so}$ is the next-nearest
neighbor spin orbit  hopping amplitude and $c_j$ is either $a_j$ or $b_j$,
depending wether the index $j$ labels an $A$ or $B$-sublattice site,
respectively. The factor $\nu _{ij}$ is $+ 1 $ if the next-nearest
neighbor hopping path rotates counterclockwise and -1 if it rotates clockwise.
%

In undoped graphene, and for $t _{so}=0$, the conduction and valence
bands touch at two inequivalent points of the Brillouin zone: ${\bf
K}=\frac 4 3 \frac {\pi} a (1,0) $ and ${\bf K}'=-\frac 4 3 \frac
{\pi} a (1,0) $, being $a$ the lattice parameter of the triangular
lattice. These are known as Dirac points. The main effect of the
SOC in the electronic spectrum is the opening of a energy gap, $6
\sqrt{3} t_{so}$, at the Dirac points. Near these points the wave
functions for each spin $s_z$  can be expressed via the ${\bf k}
\cdot {\bf P}$ approximation \cite{Ando_2005} in terms of envelope
functions $ \psi _{+,s_z} = [A _{s_z} ({\bf r}),B _{s_z} ({\bf r})]$
and $ \psi
_{-,s_z} = [A '_{s_z} ({\bf r}),B'  _{s_z} ({\bf r})]$ for
states near the ${\bf K}$  and ${\bf K}'$ points respectively.
%
These wavefunctions satisfy the Dirac equations $H_{\tau_z,s_z} \psi
_{\tau_z,s_z}= \varepsilon \psi _{\tau_z,s_z}$, where $\tau_z =\pm
1$ specifies the Dirac points ${\bf K}$ and ${\bf K}'$ and
\begin{equation}
H_{\tau_z,s _z}= v_F ( p_x \tau _z \sigma _x+ p_y \sigma _y)+ \Delta
_{so} \sigma _z \tau_z s_z. \label{HDirac}
\end{equation}
Here $v_F$=$\frac {\sqrt{3}} 2 \gamma_0 a $, $\Delta _{so} =3 \sqrt{3}t
_{so}$ and $\sigma _i$ are Pauli matrices representing  the pseudospin
degree of freedom corresponding to the two sites per unit cell of
the graphene lattice.  Note that $\vec{p}$ denotes the distance in momentum from the ${\bf K}$ and ${\bf K}'$ points.
%
It is important to note that for a given $s_z$, the Dirac mass terms
in the Hamiltonian induced by SOC at ${\bf K}$ and ${\bf K}'$ have
the same magnitude but different sign.

Graphene with intrinsic SOC is a topological insulator with a finite
spin Hall conductivity. For a given $s_z$ and valley index $\tau_z$,
the spin Hall conductivity as obtained from the Kubo formula has the form
\begin{equation}
\sigma _{xy} ^{s_z,\tau _z} =2  \frac {e^2} {\Omega} \sum
_{\vec{k},i,j} \frac {\textrm{Im}  [ < \psi ^{i}_{\vec{k}} |v _y |
\psi ^{j}  _{\vec{k}}> < \psi ^{j}  _{\vec{k}} |v _x | \psi ^{i}
_{\vec{k}}>]} {(\varepsilon^{i} _{ \vec{k}}- \varepsilon ^{j}
_{\vec{k}} ) ^2 } \label{Kubo}\, ,
\end{equation}
where $j(i)$ runs over occupied (empty) states, $v_{\nu}$ is the
velocity operator in the $\nu$ direction  and the wavefunctions $
\psi ^{i}_{\vec{k}}$ and energies $\varepsilon^{i} _{ \vec{k}}$ are
obtained by diagonalizing Eq. (\ref{HDirac}) for the corresponding
$\tau_z$ and $s_z$. In undoped graphene (Fermi energy crossing the Dirac points), the Hall conductivity
\emph{for each valley} takes the value
\begin{equation}
\sigma _{xy} ^{s_z,\tau _z} = - \frac 1 2 \frac { e^2} h s_z,
\end{equation}
where $h$ is the Planck constant. When summing over spins the total Hall conductivity of the system is
zero, as it should be for a system with time reversal symmetry \cite{Hasan_2010,Qi_2010}.
However, for each spin the Hall conductivity is quantized, although
with opposite sign. This is the signature of the quantum spin Hall
effect. Note also that although for an isolated valley the Hall
conductivity is a half integer in units of $e^2 /h$, the sum of the
${\bf K}$ and ${\bf K}'$ conductivities  is quantized to integer multiples of $e^2/h$, as it should
be  for a filled band of noninteracting
electrons \cite{Thouless_1982,Sinitsyn_2006}. In an insulator, the
value of the Hall conductivity in units of $ e^2 /h$ is related
to the first Chern number \cite{Thouless_1982} of its bandstructure. The Chern number corresponding to a  2$\times$2 Hamiltonian
\begin{equation}
H_{\tau_z,s _z}=  \mathcal{\epsilon} (\vec p) \, \, \vec {h} (\vec p) \cdot
\vec{\sigma }\, \, , \label{vech}
\end{equation}
is related to the number of times the unit sphere is covered by the
unit vector ${\vec h} (\vec p)$ when $\vec{p}$ runs over the whole
reciprocal space \cite{Rajaraman_book,Hasan_2010,Qi_2010}. This
number takes the form
\begin{equation}
n = \frac 1 {4 \pi} \int d ^2 p (\partial _{ p_x} \vec h \times
\partial _{p _y}\vec h) \cdot \vec h \label{Chern},
\end{equation}
where ${\vec h}  (\vec p)$  depends on $\tau _z$ and $s_z$ and
$\pm\mathcal {\epsilon} (\vec p)$ are the energy eigenvalues.
$n$ is a topological invariant, the Pontyagin index of the
mapping $h(\vec{p})$ \cite{Rajaraman_book}.
Physically, the vector $\vec{h}(\vec{p})$ represents
the expectation value of the orientation of the pseudospin
associated with the wavefunctions of the Hamiltonian, Eq. (\ref{HDirac}).

\begin{figure}
 \includegraphics[clip,width=9cm]{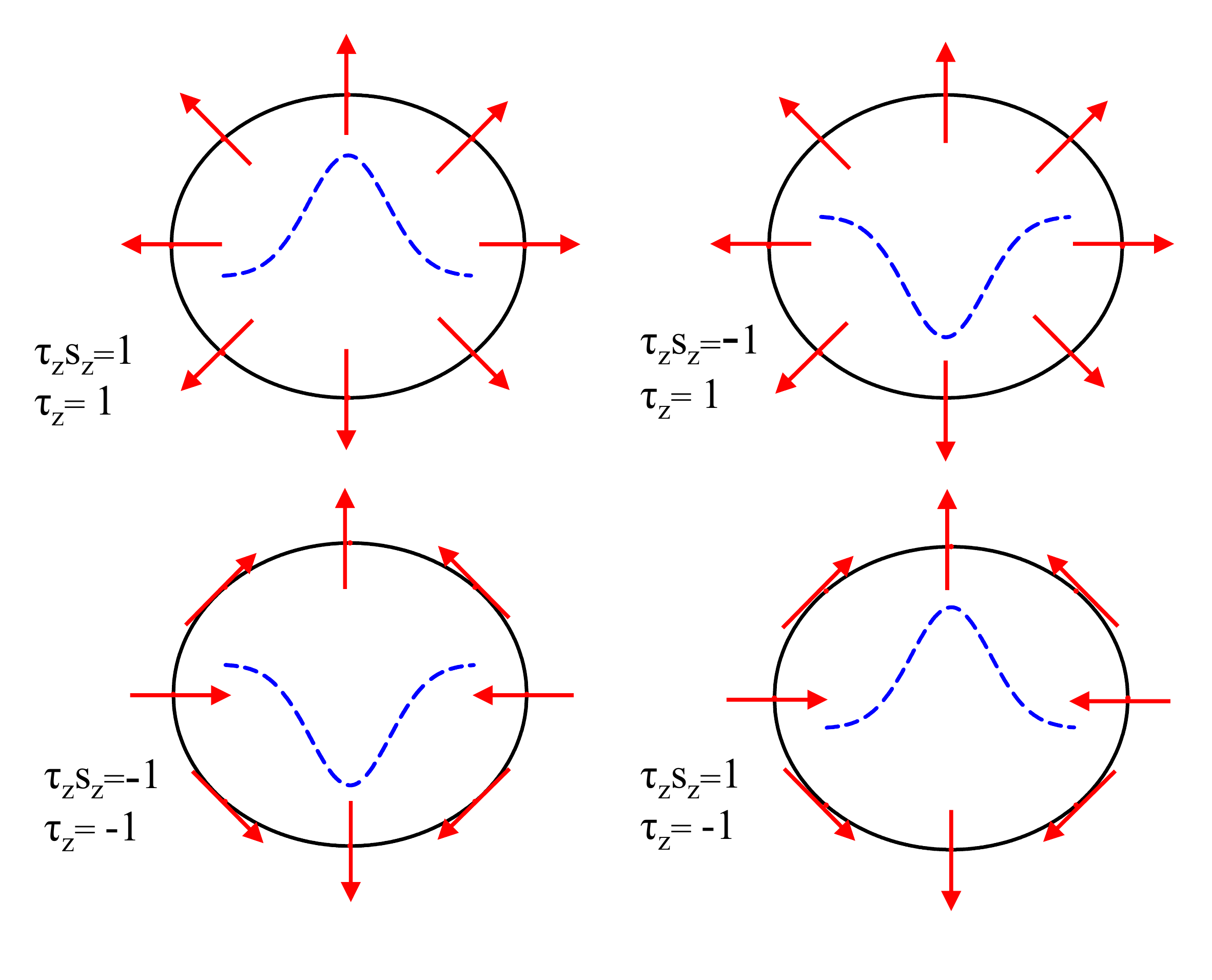}
 \caption{(Color online) Four flavors of merons, see
 Ref. \onlinecite{Girvin_book}. These are vortices that are right or left handed
 depending on the  valley index $\tau _z$. The topological charge, $\pm 1/2$, is determined by the sign of the spin $s_z$.}
  \label{merons}
\end{figure}

The vector
$\vec h$ defines the topology of the band structure and may  be
written in the form
\begin{equation}
\vec h = \left [\tau _z \sqrt {1-[h_z (p)] ^2} \cos \theta, \sqrt
{1-[h_z (p)] ^2}\sin \theta , h_z (p) \right ], \label{h_meron}
\end{equation}
with $h_z (p) = s_z \tau_z\Delta _{so}/\sqrt{v_F ^2 p ^2 + \Delta
_{so} ^2}$. Here the valley index $\tau _z= \pm $ defines the right
and left vorticity of the topological structure and $\theta$ is the
azimuthal angle made by the momentum vector $\vec p$. At
asymptotically large momentum, $h_z$ vanishes, while in the cortex
core we have $h_z =\tau _z s _z$. This implies that there are four
flavors of topological objects which are usually referred to as
merons \cite{Girvin_book,Brey_1996}, since they are essentially half
skyrmions. The four possible merons are illustrated in Fig.
\ref{merons}.  The Chern number index corresponding to the field
$\vec h$ takes the form
\begin{equation}
n= \frac {\tau _z} {4 \pi} \int _0 ^{\infty}  d ^2 p \frac 1  p
\frac {d h_z }{ d p} = - \tau _z \frac 1 2 h_z (0) = - \frac 1 2
s_z \, .
\end{equation}
Thus, the topological charge corresponding to each Dirac point is
$\pm 1/2$ depending on the sign of the electron spin.  When summing
over the contribution from both Dirac points the Chern number in
graphene is $\pm 1$. This result follows from the fact that,
topologically, a meron has half the winding number of a skyrmion,
that is, the topological object formed by the two Dirac points
together.

\subsection{Helical Edge States}

The most spectacular consequence of the existence of an insulator
with a topologically non trivial band structure is the appearance of
gapless conducting states at interfaces where the topology of the
band structure changes. In  two dimensional quantum spin Hall
systems, helical edge states appear at the surfaces of the material.
At each edge  there exists a pair of one dimensional channels with
opposite spins that propagate in opposite directions. In
graphene these states have been obtained numerically by
diagonalizing nanoribbons terminated with different geometries,
zigzag\cite{Kane_2005b,Li_2010} or armchair\cite{Zarea_2007}. Here
we describe analytically how helical one dimensional  channels
appear at an armchair terminated edge.

\begin{figure}
 \includegraphics[clip,width=9cm]{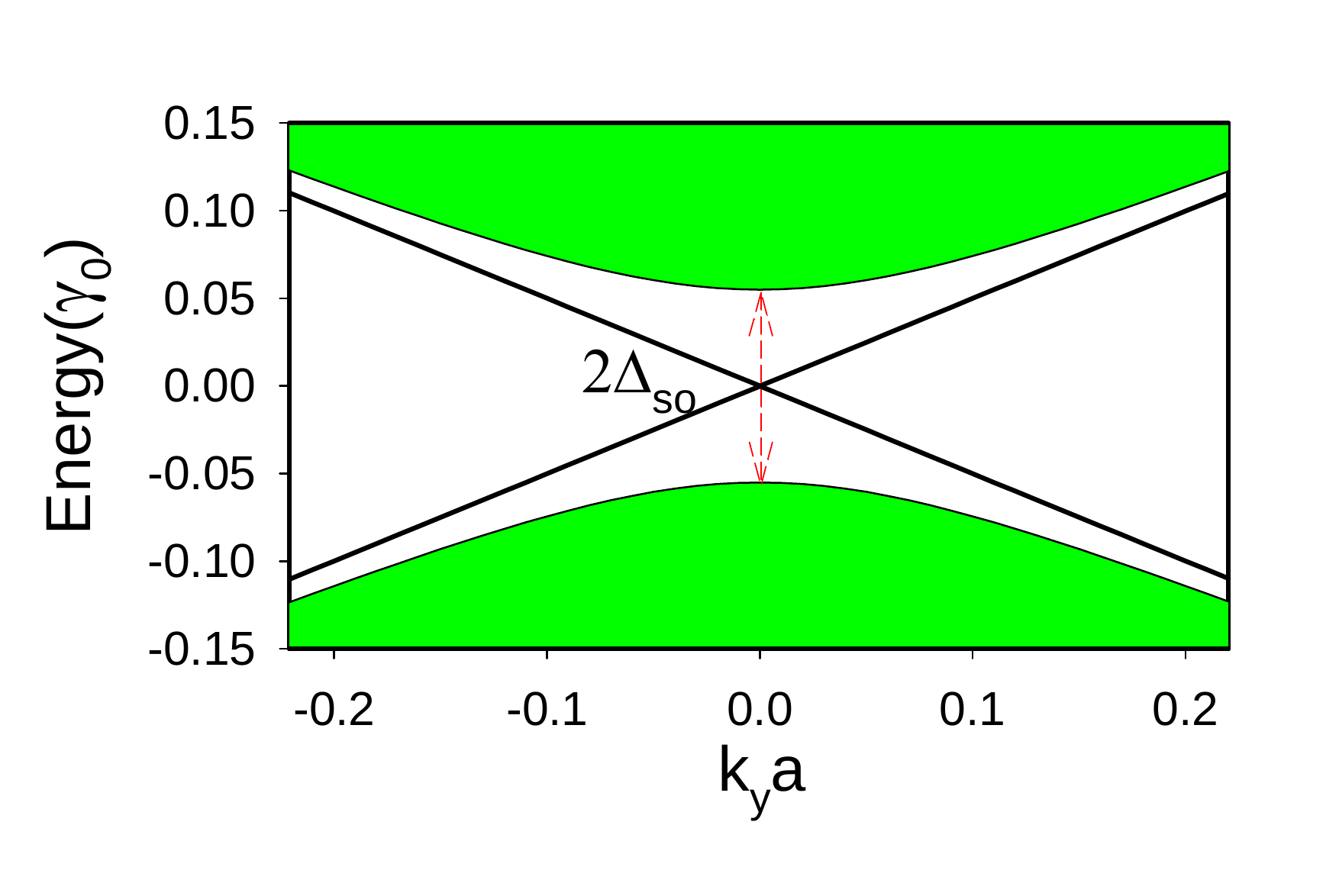}
 \caption{(Color online) Tight-binding band structure of an armchair terminated wide nanoribbon. Shadow regions represent bulk states.}
  \label{Monolayer_AC_SO}
\end{figure}

We consider an edge running along the $\hat y $-direction and the
vacuum region is defined by the $x<0$ condition. The graphene
armchair termination consists of a line of $A$-$B$ dimers, so it is
natural to have the wave function amplitude vanish on both
sublattices at $x=0$. To do this we must admix
valleys\cite{Brey_2006a} and require
\begin{eqnarray}
A_{s_z} (x=0)+A'_{s_z}(x=0) & = & 0 \, \, \, \, \, \textrm{and } \, \, \, \nonumber \\
B_{s_z} (x=0)+B'_{s_z}(x=0) & = & 0. \label{AC_BC}
\end{eqnarray}
Note that the boundary conditions should be satisfied by each spin
separately. We consider solutions with momentum $p_y$ along the
surface and energy $E$ inside the gap, $|E|<\Delta_{so}$, where $p_x=i\hbar \kappa$. The
general solution has the form
\begin{equation}\label{mono_wf}
\left [ \alpha \left ( \begin{array}{c}
        \sin \xi \\
        \cos \xi
      \end{array} \right ) e ^{i K_x x} + \beta \left ( \begin{array}{c}
        \sin \xi ' \\
        \cos \xi '
      \end{array} \right ) e ^{i K '_x x} \right ] e^ {i \frac{p_y}{\hbar}y} e
      ^{ -\kappa x }
\end{equation}
with $\hbar v _F \kappa = \sqrt { \Delta _{so} ^2 - E ^2 + v_F ^2  p_y ^2}$ and
\begin{equation}\label{tangentes}
\tan \xi = i  \frac {v _ F (\hbar\kappa - p_y)} {E - \Delta_{so} s_z } \, \, \,
\,\textrm{ and} \, \, \, \,\tan \xi' = -i \frac {v _ F (\hbar\kappa+p_y)} {E +
\Delta_{so} s_z }.
\end{equation}
The boundary conditions, Eq. (\ref{AC_BC}), imply that $\tan \xi =
\tan \xi '$. This, together with the definition of $\kappa$, gives
the solutions
\begin{equation}\label{helicoidal}
E = s_z v _ F p _y \, \, \, \, \, \, \textrm{and} \, \, \, \, \, \, \hbar v_F \kappa = \Delta _{so}
\, \, \, .
\end{equation}
Since the graphene flake exists in the $x>0$ region, normalizability of the wavefunction implies that $\kappa$ should be positive for any spin. Therefore,
quasiparticles with positive spin, $s_z>0$, move in the positive $\hat y$ direction and have and energy $E=v_ F p _y $. On the contrary,
quasiparticles with $s_z <0$ move in the negative $\hat y$ direction and have energy $ E = - v _F p _y $. On the opposite surface, for which the wavefunction exists for $x<0$, normalizability implies that $p_x=-i\hbar\kappa$. In this case $E=-s_zv_Fp_y$, and the edge states with positive (negative) velocity in the $\hat y$-direction have $s_z <0$ ($s_z >0$).
In Fig. \ref{Monolayer_AC_SO} we plot the band structure of a wide armchair terminated nanoribbon with a SOC $t _{so} =0.01\gamma_0$.

\section{Graphene bilayers}
Bilayer graphene consists of two stacked graphene sheets. In bilayer
graphene there are four sites per unit cell, which we label $A_1$, $B_1$ and $A_2$ and $B_2$  in the the first and second
layer respectively. We consider the so-called Bernal stacking, commonly found experimentally, in which site $B_2$ is exactly on top of the sublattice $A_1$. Interlayer coupling is modeled by hopping amplitude $\gamma_1$ between these two sites in each unit cell.
The low energy properties of this model are well described for each spin $s_z$
by the following Hamiltonian acting on the four-component spinor $(A_1,B_1,A_2,B_2)$ \cite{McCann_2006},
\begin{equation}
H^{BG}_{\tau_z,s_z} = \mathcal{T}_0\otimes H_{\tau_z,s_z}- \frac {\gamma _1} 2 \left
(\mathcal{T} _x \otimes \sigma _x - \mathcal{T}_y \otimes \sigma _y
\right )\, \, \, , \label{Dirac_bilayer}
\end{equation}
where $H_{\tau_z,s_z}$ is the monolayer Hamiltonian of Eq. (\ref{HDirac}), $\sigma _i$  and $\mathcal{T}_i$ are the Pauli matrices  for
the sublattice  and layer degree of freedom respectively and
$\mathcal{T} _0$ is  the unit matrix in the layer  subspace.
The four energy bands of $H^{BG}$, denoted by $\varepsilon _{\pm} ^{(1,2)} (p)$, are
\begin{equation}
\varepsilon _{\pm} ^{(\alpha)}=\pm \sqrt{v_ F ^2 p ^2 + \Delta _{so} ^2 +
\frac {\gamma_1 ^2 }  2 + (-1) ^{\alpha} \sqrt {v_F ^2 p ^2 \gamma _1 ^2 + \frac {\gamma _1 ^4} 4 }}.
\end{equation}
These energies are independent of $s_z$ and $\tau_z$ \cite{nota_elsa}.
The eigenvalues $\varepsilon _{\pm} ^{(2)}$ describe two strong
interlayer coupling bands with energies $\varepsilon _  + ^{(2)}
\geq \gamma _1$ and $\varepsilon _ - ^{(2)} \leq -\gamma _1$.  These
bands do not touch at the Dirac points, and correspond to
wavefunctions mostly localized at sites  $A_1$ and $B_2$, which form
strong dimers \cite{McCann_SSC}. The eigenvalues $\varepsilon _{\pm}
^{(1)}$ describe low energy bands. Performing the usual low energy
approximation $v_Fp\ll\gamma_1$, the dispersion of these bands can
be approximated by $\varepsilon _{\pm} ^{(1)} = \pm \sqrt { \frac
{p^4}{ 4 m ^2} + \Delta _{so} ^2 }$, where $m$ is an effective mass
induced by the interlayer hopping, $m= \gamma _1 /2v_F ^2$. The
corresponding low energy eigenstates are mostly localized at sites
$B_1$ and $A_2$. All of these states are degenerate in the spin
$s_z$ and valley $\tau _z$ indices.

For a given $s_z$ and $\tau _z$, the Hall
conductivity can be obtained numerically by plugging the eigenvalues
and eigenfunctions  of the Hamiltonian (\ref{Dirac_bilayer}) into the
expression (\ref{Kubo}). For charge-neutral bilayer graphene, only negative energy bands are filled, and the Hall
conductivity takes the form
\begin{equation}
\sigma _{xy} ^{s_z,\tau _z} = - \frac { e^2} h s_z.
\end{equation}
When summing over the valleys, the Hall conductivity per spin is
twice that of graphene. Interestingly, this is the same result one would
expect for the case of two decoupled layers, although the eigenfunctions are completely different from those of the coupled bilayer. As in the case of a
monolayer, time reversal symmetry dictates opposite $\sigma _{xy}^{s_z,\tau _z}$ upon reversing the spin, so that the total Hall conductivity of the system is zero.

\begin{figure}
 \includegraphics[clip,width=9cm]{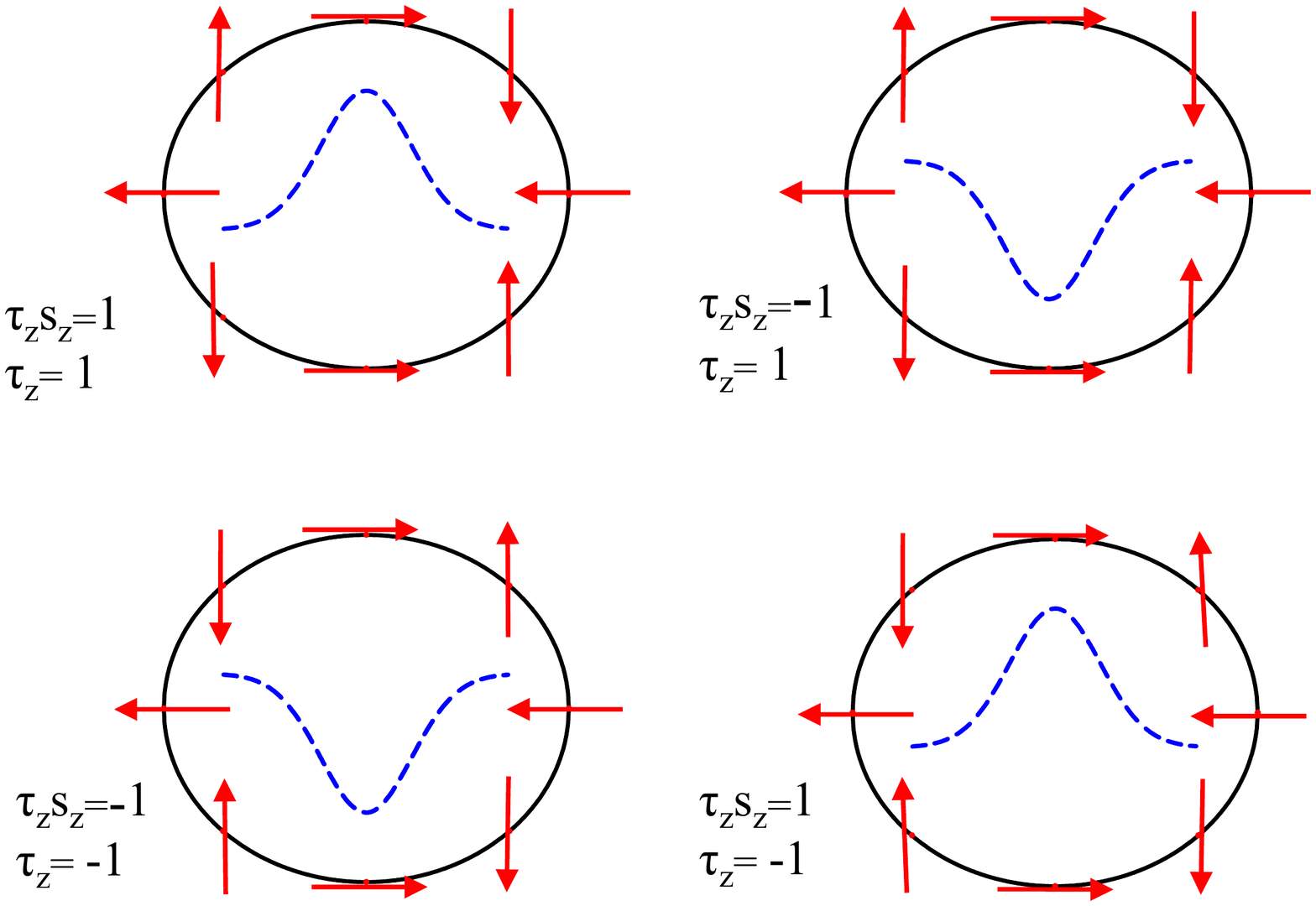}
 \caption{(Color online) Four flavors of double
vortex merons. These are vortices that are right or left handed
 depending on the  valley index $\tau _z$. The topological charge $\pm 1$ is determined by the sign of the spin.}
  \label{bimerons}
\end{figure}

\subsection{Band structure topology}
The value of the Hall conductivity can be understood from the
topology of the band structure through the first Chern number. Because of the gap $\sim 2\gamma_1$ between the high energy dimer bands $\varepsilon_{\pm}^{(2)}$, these cannot change their topology by adding a weak SOC. Therefore, only the gapless $\varepsilon_{\pm}^{(1)}$ bands acquire non-trivial topology that contributes to a finite Chern number $n$. To compute $n$, it is then sufficient to consider the effective 2$\times$2 low energy Hamiltonian that
describes the low energy bands of the system in the limit $\varepsilon\ll \gamma_1$ \cite{McCann_2006},
\begin{eqnarray}
H _{\tau _z,s_z} \! \!  \! &  =  &  \! \!  \! \left(
                   \begin{array}{cc}
                     \Delta_{so} \tau _z s_z  & -\frac {(\tau_z p_x -i p _y) ^2}{2m} \\
                     - \frac {(\tau_z p_x + i p_y)^2}  {2m}& -\Delta _{so} \tau _z s _z \\
                   \end{array}
                 \right) \nonumber \\
                 & = & \Delta _{so} \tau_z s_z \sigma _z- \frac
                  {(p_x ^2-p_y ^2)} {2 m}\sigma _x - \frac  { p_x
                 p_y} {m} \tau_z\sigma _y \label{DiracBG-effective}\, \, .
\end{eqnarray}
This Hamiltonian acts on the two-component spinor $(B_1,A_2)$. This Hamiltonian describes the low energy effective
coupling between carbon atoms in different layers which are not
directly connected by tunneling. Their coupling arises as a result of virtual transitions through the high energy dimer states that have been integrated out.  From this Hamiltonian  and the
relation of Eq. (\ref{vech}) we get the following expression for  the unit
vector field $\vec{h} (\vec{p})$:
\begin{equation}
\vec h  = \left [- \sqrt {1-[h_z ] ^2} \cos 2\theta, -\tau _z \sqrt
{1-[h_z ] ^2}\sin 2\theta , h_z ) \right ], \label{h_bimeron}
\end{equation}
with $h_z (\vec{p})= \Delta _{so} \tau_z s_z / \sqrt{ \Delta _{so} ^2 +
\left (\frac { p^2 } { 2 m} \right ) ^2}$. As in the case of a
monolayer, the valley index $\tau _z$ determines the vorticity of the field $\vec{h}(\vec{p})$ and the
angle $\theta$  is the azimuthal angle of momentum
$\vec{p}$. At large momentum $p$, the $\hat{z}$-component of the
field $\vec{h}$ vanishes and the field is confined to be in the
$x-y$ plane.  As in the case of the monolayer, there are four flavors
of these topological objects, shown in Fig. \ref{bimerons}. The Chern number of the corresponding fields $\vec h(\vec p)$ is
\begin{equation}
n= 2 \frac {\tau _z} {4 \pi} \int _0 ^{\infty}  d ^2 p \frac 1  p
\frac {d h_z }{ d p} = - \tau _z  h_z (0) = -  s_z \label{ChernBG} .
\end{equation}
Thus, the Chern number corresponding to each valley is $\pm 1$
depending on the sign of the electron spin. It is interesting to
note that, although these objects have twice the Pontryagin charge
of merons (which are often thought as half a skyrmions), they are
not skyrmions. Far away from the center of a skyrmion, i.e., at  $p
\rightarrow \infty$,  the orientation of the vector field $\vec{h}
(\vec{p})$  is constant. This is not the case of bilayer graphene
(see Fig. \ref{bimerons}), although, as in a skyrmion, the Chern
number is unity for each $\tau_z,s_z$. We  call these objects double
vortex merons.
When summing over valleys the Chern number
in bilayer graphene is $\pm 2$ (depending on the spin), in agreement with the results
obtained for the Hall conductivity.

\subsection{Edge states}
\begin{figure}
 \includegraphics[clip,width=9cm]{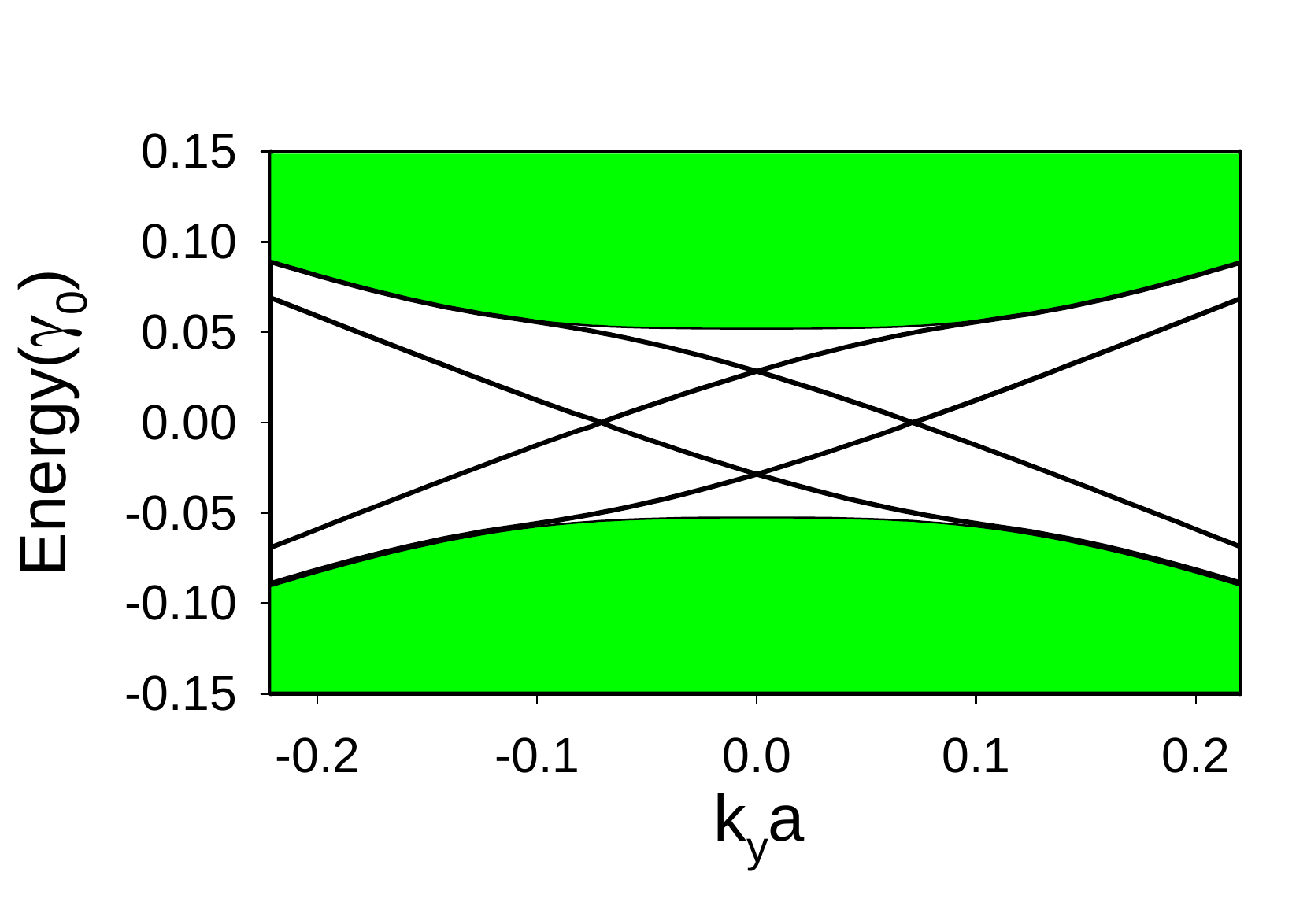}
 \caption{(Color online) Band structure of a wide armchair terminated bilayer graphene ribbon
 for a single spin. The (green) shaded areas represent the continuum bulk states.}
  \label{Bilayer_AC_SO}
\end{figure}

According to the bulk-boundary correspondence rule
\cite{Hasan_2010}, and since the Chern number of bilayer graphene is
twice that of the monolayer, the number of helical surface states
should also be  double. In order to analyze the edge states, we
diagonalize the tight-binding Hamiltonian corresponding to a  wide
armchair terminated bilayer graphene ribbon. In Fig.
\ref{Bilayer_AC_SO} we plot its band structure as function of the
wavevector $k_y$ along the ribbon for a given electron spin $s_z$.
As in the case of a monolayer of graphene, the Hamiltonian commutes
with $s_z$ and the band structure is degenerate in spin. The bulk
bands are indicated by shaded regions. We identify two pairs of
parallel edge states inside the gap. The states with positive
velocity are located on the opposite ribbon edge from  the states
with negative velocity. This can be seen by analyzing the
eigenfunctions of the low-energy Hamiltonian, Eq.
(\ref{DiracBG-effective}). As a result of the helicity of the edge
states, their location changes to the opposite edge when the spin of
the carriers changes sign.

Bilayer graphene edge states can be understood as
bonding/antibonding combinations of the surface states of the two
constituent graphene layers. In the absence of tunneling between the
graphene layers, the surface states have the form of
Eq. (\ref{mono_wf}) with $ \xi$, $\xi '$, $E$ and
$\kappa$ determined by Eq. (\ref{tangentes}) and Eq. (\ref{helicoidal}).
For finite $\gamma_1$, surface states from  layers 1 and 2 become
coupled.
To leading order in $\gamma_1$, the energies  and wavefunctions  of the graphene bilayer Hamiltonian Eq. (\ref{Dirac_bilayer}), defined for $x<0$, are
\begin{equation}
E^{s_z} _{\pm} = s_z   v_F p_y \mp \frac {\gamma_1} 4
\label{ene_bilayer}
\end{equation}
\begin{equation}
\psi^{s_z} _{\pm} =\frac 1 {2\sqrt{2}} \left ( \begin{array}{c}
                                                       -i s_z \\
                                                       1 \\
                                                        \pm s_z \\
                                                       \pm i
                                                     \end{array}
\right )\left [ e ^{i K_x x} - e ^{i K_x ' x} \right ] e ^{i k_y y}
e ^{-\Delta _{so} x/\hbar v_F} \, \, \, \, . \label{wf_bilayer}
\end{equation}
where $\pm$ correspond to bonding/antibonding states.
These expressions are valid when the SOC $\Delta
_{so}$ is much larger than $\gamma_1 /4$. In other situations, bulk
states should reduce the bonding/antibonding energy separation.
\begin{figure}
 \includegraphics[clip,width=8cm]{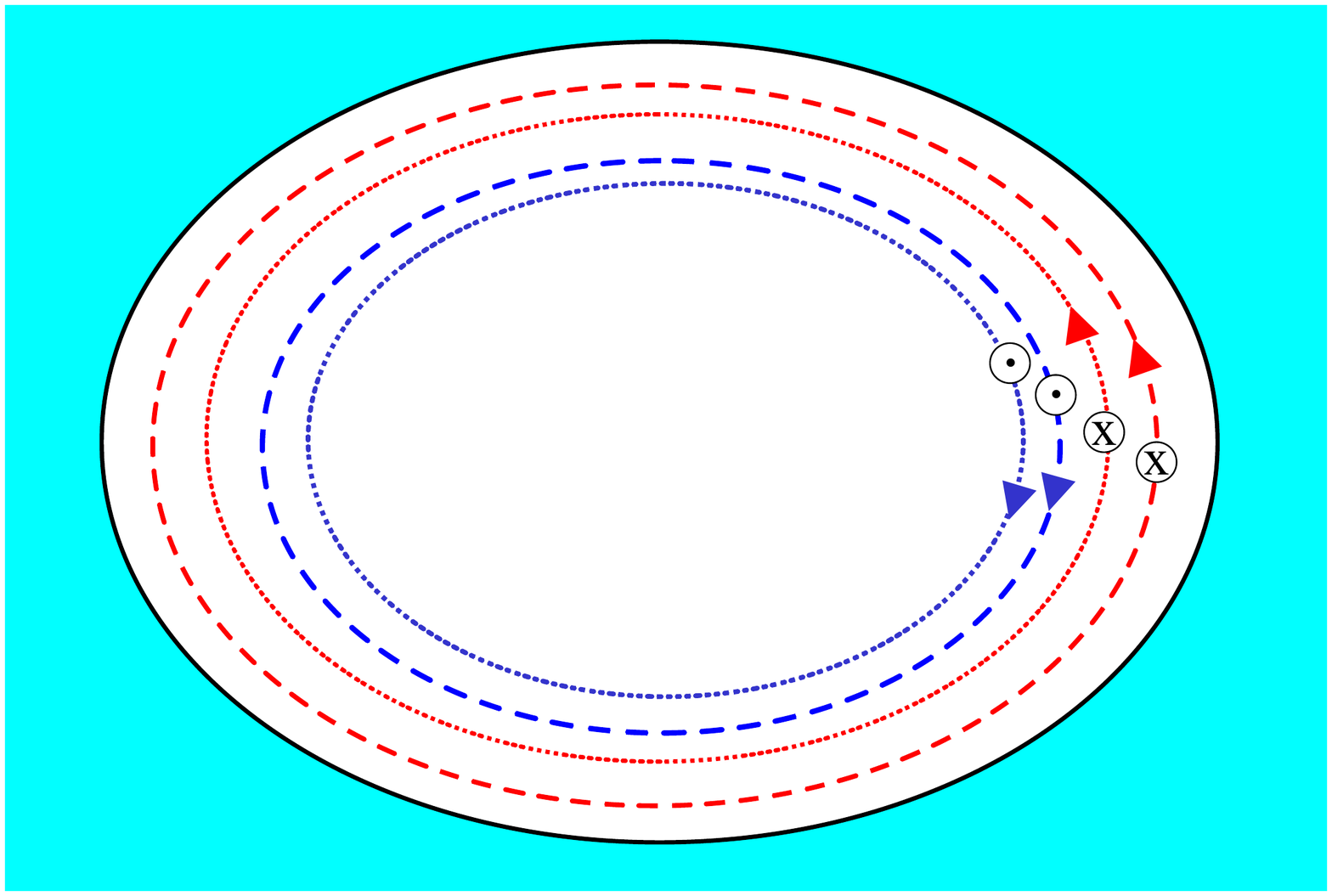}
 \caption{(Color online) Schematic diagram of the edge states in a bilayer graphene system.
 At the edge of the sample there are four channels: two with spin up and moving in one direction,
 and two with spin down an moving in the opposite direction. Dotted and dashed lines represent
 channels formed by bonding/antibonding combinations of individual layers edge channels.
 }
  \label{Scheme_channels}
\end{figure}

By inspection of Eq. (\ref{wf_bilayer}), we see that at each edge of
the system there exist four one dimensional channels: two with spin
up moving in one direction and the other two with spin down and
moving in the opposite direction, see Fig. \ref{Scheme_channels}.
The two terminal electrical conductance in this model is $4 e^2/h$.
The injection of charge current through edge states would result in
antisymmetric spin accumulation at the edges. In a four terminal
geometry the spin conductances should be quantized.

However, in bilayer graphene, backscattering between channels at the
same edge moving in opposite directions with opposite spins is not
forbidden, in contrast to the case of monolayer graphene. Indeed,
Kramers's theorem ensures degeneracy of time-reversed pairs in the
absence of time reversal symmetry breaking, preventing
backscattering between energy degenerate edge states with opposite
spins at high symmetry points of the Brillouin zone, such as the
$\Gamma$-point (for which $\vec k=-\vec{k}$ modulo a reciprocal
lattice vector) \cite{Hasan_2010}. While in monolayer graphene edge
states branches cross at the $\Gamma$-point $k_y=0$, bilayer
graphene edge state branches cross away from $k_y=0$, see Fig.
\ref{Bilayer_AC_SO}. As a result, SOC perturbations will in general
split their degeneracy, and open a gap in the edge state dispersion
around zero energy.
Therefore, this model is not, strictly speaking, a topological insulator, since there exist single-body effects that are symmetric under time reversal that can induce backscattering between edge channels, and make the edges insulating. This happens for any two-dimensional insulator with an even number $N_K$ of edge state pairs, related by time reversal symmetry ($k_y, s_z\to -k_y, -s_z$), at any given energy inside the gap ($N_K=2$ for the case at hand). An odd number of pairs $N_K$, like in monolayers or trilayers with SOC, implies that there exists at least one of those pairs crossing at a high symmetry point, and are therefore protected from scattering by Kramer's theorem.
A $\mathbb{Z} _2$ topological invariant of the form
\begin{equation}
\nu = N_K \textrm{mod } 2,
\end{equation}
is defined \cite{Hasan_2010} to differentiate strict topological
insulators (dubbed `strong'), $\nu =1$, from  `weak' topological
insulators, with $\nu =0$ and $N_K\neq 0$. Within this
classification, bilayer graphene is a weak topological insulator.
The `weakness' is in relation to the class of
time-reversal-invariant perturbations that can open a gap in the
edge states, in this case general spin-mixing perturbations.

%
\subsection{Stability of the topological insulating phase}

The non-trivial insulating phase created by the instrinsic SOC may
be destroyed by Rashba SOC or other perturbations if they are strong
enough. This happens not by inducing backscattering between edge
states, but by band reconnection, which changes the topology of the
bulk bandstructure back to that of a conventional insulator
\cite{Kane_2005a}. It is a generic possibility both in weak and
strong topological insulators.

We now consider the destruction of bilayer graphene's topological insulator phase by a symmetric Rashba SOC, a staggered sublattice mass term and a
voltage bias between the two layers. We also discuss the effect of
the trigonal warping on the topological insulating phase.
\par \noindent
1) The\emph{ Rashba spin orbit coupling} term arises due to an
electric field perpendicular to the bilayer plane or from the
interaction with a substrate,
\begin{equation}
H_R = \lambda _R \mathcal{T} _0 \otimes \left (\tau _z \sigma _x
\otimes s _y- \sigma _y \otimes s_x \right ). \label{HRashba}
\end{equation}
Here $s_x$ and $s_y$ are the Pauli matrices in the spin subspace and
$\lambda _R$ is the intensity of the coupling. This term violates
mirror symmetry about the planes. In the presence of this term the
$z$-component of the electron spin is not conserved, but  there is a
region of values of the coupling, $0< \lambda _R < \Delta _{so}$,
for which the ground state is adiabatically
connected\cite{Kane_2005a,Kane_2005b} to the topological insulating
phase. For values of the coupling $\lambda _R
> \Delta _{so}$ the energy gap closes and the electronic structure
is that of a zero gap semiconductor with quadratically dispersive
bands.
\par \noindent
2) A \emph{staggered sublattice potential} of the form
\begin{equation}
H_v = \lambda _v \mathcal{T} _0 \otimes \sigma _z
\end{equation}
stabilizes the band insulator phase that competes with the
topological insulating phase. This term is typically zero in graphene but
would be present for a similar boron nitride film. In the effective
2$\times$2 Hamiltonian this term appears as a mass term independent
of the spin and valley, $\lambda _v\sigma _z$. In the presence of this term, the form of the unit
vector field $\vec{h} (\vec{p})$ of Eq. (\ref{h_bimeron}) is unaffected,
but now the $z$-component takes the form
\begin{equation}
h_z (p)= \frac {\Delta _{so} \tau_z s_z  + \lambda _v } { \sqrt{ (
\Delta _{so} \tau_z s _z + \lambda _v )  ^2 + \left (\frac { p^2 } {
2 m} \right ) ^2} } \, \, .
\end{equation}
The Chern number associated with this vector field $\vec{h}
(\vec{p})$ is
\begin{equation}
n = -\tau _z \, \textrm{sign} (\Delta _{so} \tau_z s_z  + \lambda _v
)
\end{equation}
For values  $\lambda _v < \Delta _{so}$ the Chern number is $-s _z$ (equal for both valleys) and the system is a topological insulator. For values $\lambda _ v>\Delta _{so}$ we have $n = - \tau _z$ and the  total Chern number per spin is zero when summing over
valleys. In this case the bilayer graphene becomes a  band  insulator.
\par \noindent
3) A \emph{bias voltage} between the layers gives a contribution to
the bilayer Hamiltonian, Eq. (\ref{Dirac_bilayer}), of the form
\begin{equation}
H_{bias} = V \mathcal{ T} _z \otimes \sigma _0.
\end{equation}
This term acts equally on both sublattices and on both spin
orientations. In the effective 2$\times$2 low energy Hamiltonian Eq. (\ref{DiracBG-effective}), this term has
exactly the same form as the staggered sublattice potential $ V
\sigma _z$. Thus, the condition for the existence of a topological
insulating phase in the bilayer is $V <  \Delta _{so}$.

In the last two cases, when $\lambda _v>\Delta _{so}$ or $V > \Delta _{so}$, the
system becomes a conventional insulator with vanishing Hall and spin Hall conductances. The two inequivalent valleys, considered independently, have
nonzero topological charge with opposite sign, $n = - \tau _z$, and
a valley dependent Hall effect may occur \cite{Di_2007}. The nonzero
Chern number of each valley induces chiral modes in topological domain walls
in graphene \cite{Wang_2009} and bilayer graphene \cite{Martin_2008}.

\par
\begin{figure}
\includegraphics[clip,width=8cm]{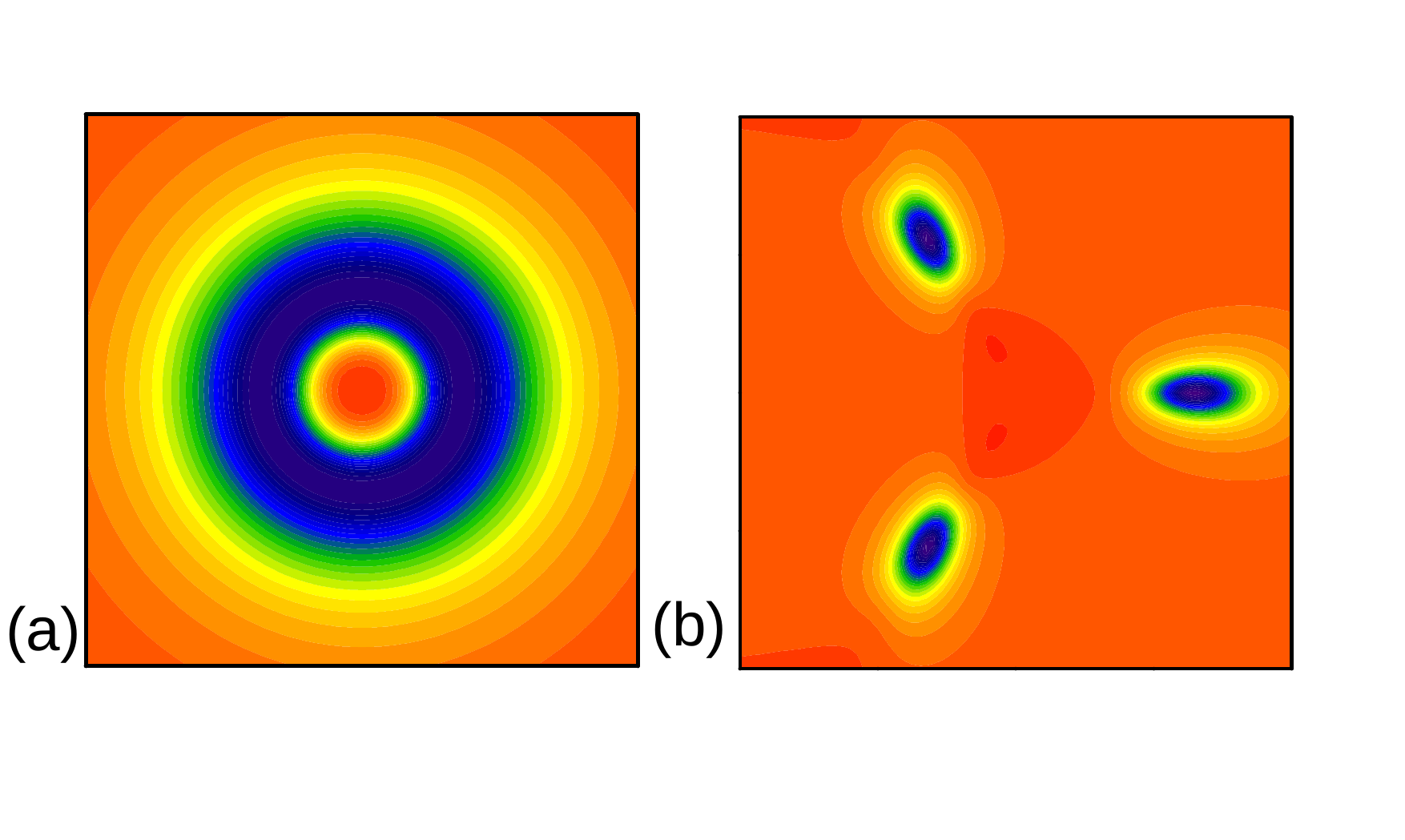}
\caption{(Color online) Density plot in the reciprocal space of the integrand of
Eq. (\ref{ChernBG}) in the absence (a) and the presence (b) of trigonal warping. The
center of each square corresponds to the original Dirac point at $\mathbf{K}$.
Cold colors represent regions where the integrand is
bigger. The parameters used in these plots are $\gamma_1$=0.1$\gamma_0$,
$\Delta_{so}=10^{-5}\gamma_0$ and in
(b) $\gamma_3$=0.05$\gamma_0$.}
 \label{Warping}
\end{figure}

 \noindent
4) \emph{Trigonal warping.} The inclusion in the Hamiltonian of
bilayer graphene of a weak direct $A_2$-$B_1$ hopping term $\gamma_3
\ll\gamma _1$  introduces  in the 2$\times$2 effective Hamiltonian a
term of the form
\begin{equation}
H _w = v_3 (\tau _z p _x \sigma _x - p_y \sigma _y), \label{Hwarping}
\end{equation}
with $v_3 = \frac {\sqrt{3}} 2 a \gamma _3/\hbar$. This term produces a
trigonal warping in the band structure, which stretches the isoenergy
lines along the directions $\phi =0$, $\frac 2 3 \pi$ and $\frac 4 3
\pi$ for the valley $\mathbf{K}$ and along the directions $\phi=\frac 1 3 \pi$, $\pi$ and $\frac 5 3 \pi$ for the valley
$\mathbf{K}'$\cite{McCann_SSC}. For $\Delta _{so}$=0, the trigonal
warping produces a dramatic change in the band structure. For a given valley, instead of
two parabolic bands touching at the Dirac point, there are now four Dirac points,
one at the center ($\mathbf{K}$ or $\mathbf{K'}$-point), and three others that occurs at finite
momentum and in the directions $\phi$ mentioned above.  The SOC
opens gaps at the four Dirac points, making the system insulating. To
analyze the nature of the insulating phase we compute the Chern
number. The warping term breaks the symmetric form of the unit vector field $\vec{h} (\vec{p})$ given in Eq. (\ref{h_bimeron}) and
prevents an analytical calculation of the Chern number. By
integrating numerically Eq. (\ref{Kubo}) we obtain that the Chern
number is not affected by the trigonal warping.  In the absence
of trigonal warping the main contribution to the integral
comes from an annulus  around each Dirac point where $dh_z/dp$ is
maximum, see Fig. \ref{Warping}(a) (dark region).
When the trigonal warping effects are taken into account, the
main contribution to the Chern number comes from the regions around the
three new Dirac points, Fig. \ref{Warping}(b).

\section{Spin-orbit proximity effect}
In this Section we analyze the case of a graphene bilayer in which only one
of the layers has a finite intrinsic SOC $\Lambda_{so}$, whereas the other
layer has zero coupling. This could happen in a bilayer in contact
with a strong SOC metal\cite{Dedkov_2008}, or in a
bilayer where heavy atoms or molecules have been deposited on top of
one of the two layers\cite{Castro_Neto_2009}.

\begin{figure}
\includegraphics[clip,width=8cm]{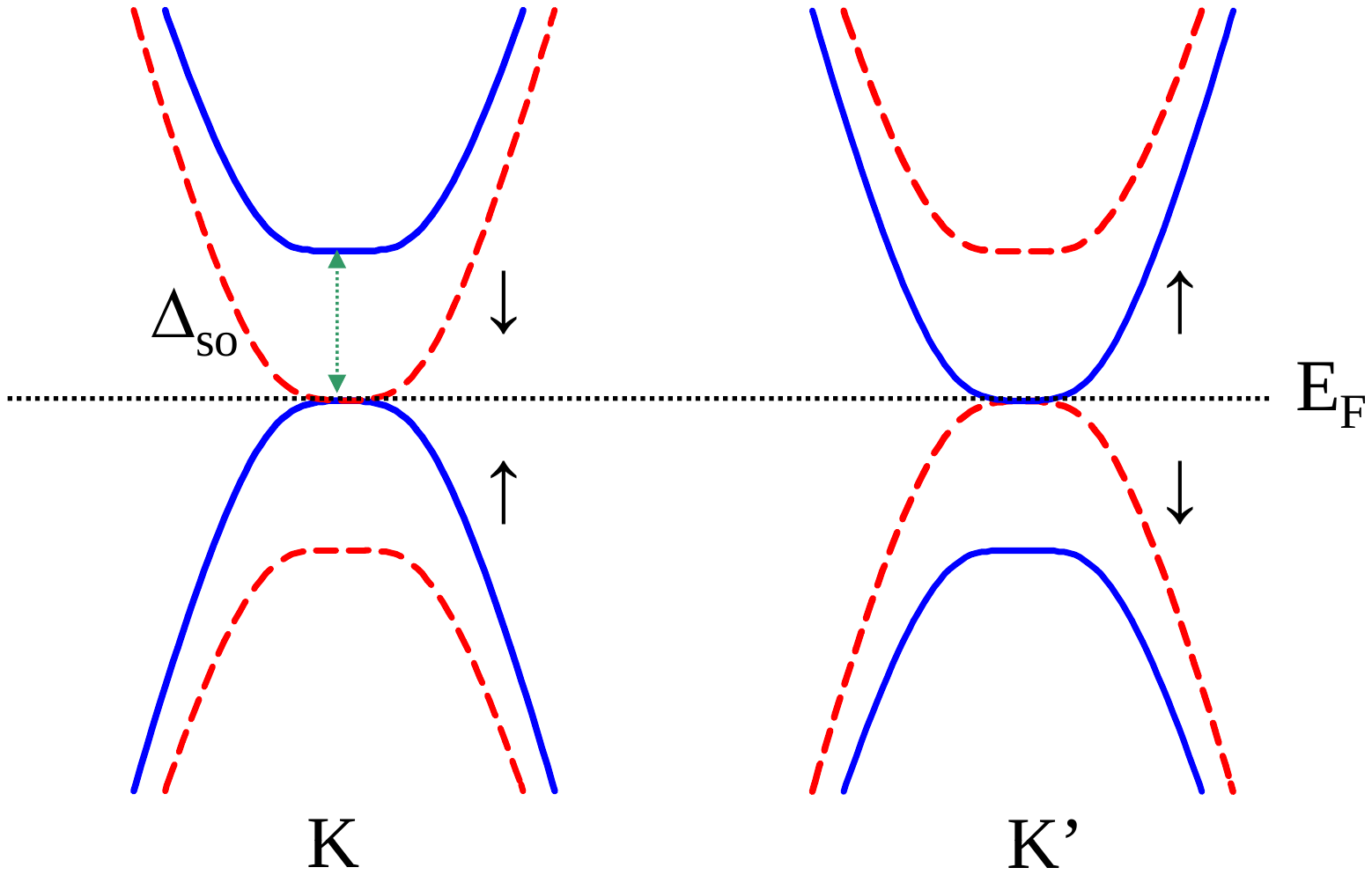}
\caption{(Color online) Low energy band structure of a graphene bilayer where only one of the two layers
has an finite SOC $\Delta _{so}$. Left and right panels correspond to the bands near the
$\mathbf{K}$ and $\mathbf{K}'$ points, respectively. Bands for states with spin up $s_z=1$ (down $s_z=-1$) are represented with solid (dashed) lines. The Fermi energy is indicated by an dotted horizontal line. The SOC opens a gap $\Delta _{so}$ with a different sign for each valley and each spin orientation.}
 \label{Bandas_proxi}
\end{figure}

\begin{figure}
\includegraphics[clip,width=8cm]{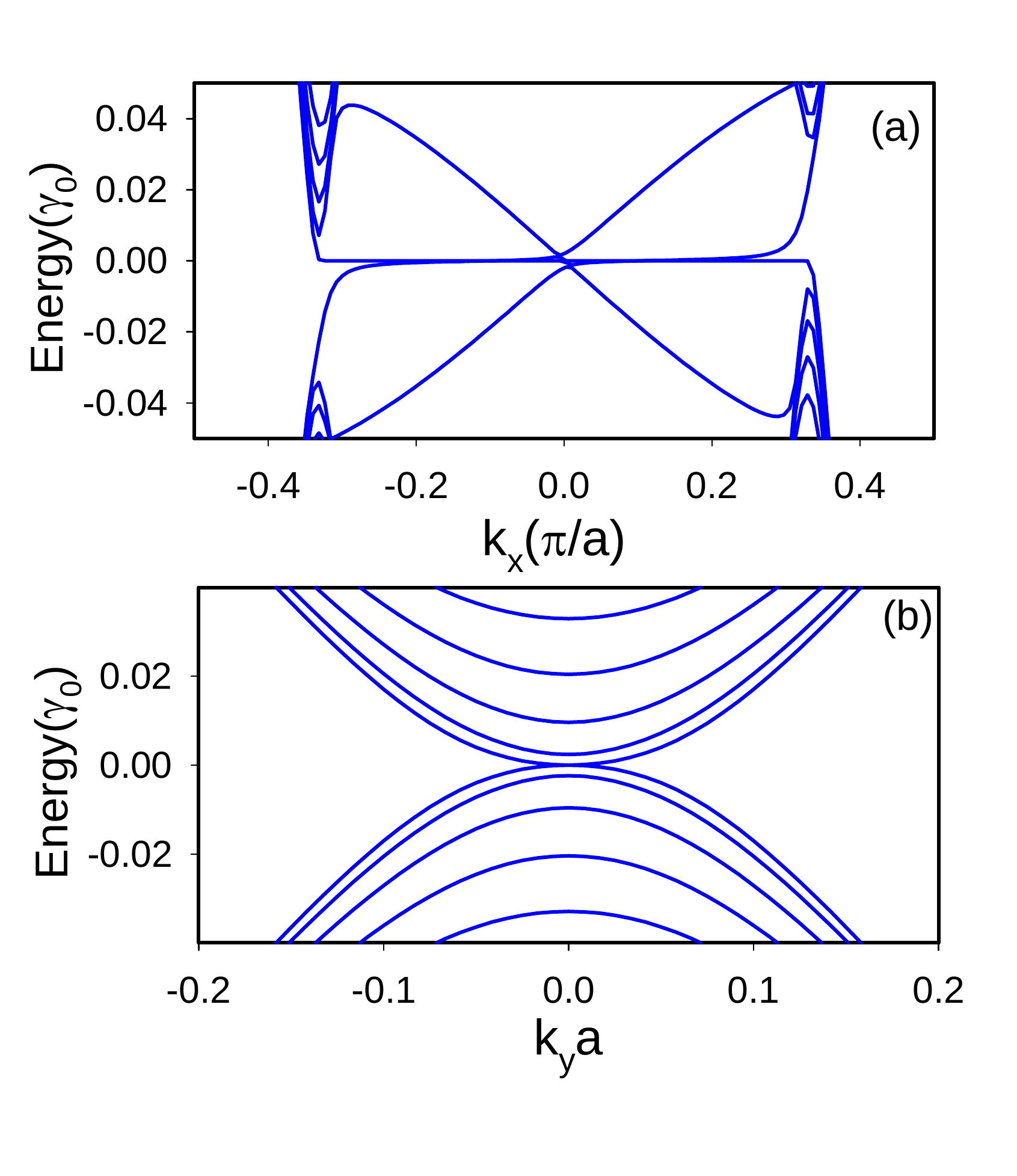}
\caption{(Color online) Tight binding band structure of a zigzag (a) and a armchair (b) terminated
bilayer graphene nanoribbon in which there is SOC only in one of the layers.
 These results correspond to $s_z$=1. Results for $s_z$=-1 are obtained by changing $\vec{k}$ to $-\vec{k}$.
In the numerics $\gamma_1 =0.1\gamma_0$ and $\Delta _{so} =3 \sqrt{3} \times 10^{-2} \gamma_0$.
The width of the armchair and zigzag nanoribbons is 151$a$ and 120$\sqrt{3}a$ respectively.
In the armchair case all the states are extended across the ribbon. In the zigzag one the states in the center of the Brillouin zone are localized at the edges. }
 \label{edge_proxi}
\end{figure}

When only one of the layers is affected by the SOC,
the 2$\times$2 effective Hamiltonian takes the form
\begin{eqnarray}
&&H _{\tau _z,s_z} =   \left(
                  \begin{array}{cc}
                    \Delta_{so} \tau _z s_z  & -\frac {(\tau _z p_x -i p _y) ^2}{2m} \\
                    - \frac {(\tau _z p_x + i p_y)^2}  {2m}& 0 \\
                  \end{array}
                \right) \nonumber \\
                & = &  \frac {\Delta _{so}} 2 \tau_z s_z (\sigma _0 + \sigma _z)- \frac
                 {(p_x ^2-p_y ^2)} {2 m}\sigma _x - \frac  { \tau _zp_x
                p_y} {m} \sigma _y
                \label{HProxi}
\end{eqnarray}
and the unit vector field $ \vec{h}$ has almost the same form as in
Eq. (\ref{h_bimeron}). Thus, the Chern number of this system is equal
to that of a graphene bilayer with intrinsic SOC in both layers,
\begin{equation}
n=  - s_z \, .
\end{equation}
Although from the value of the Chern number the system appears to be
a topological insulator, this is not the case: it turns out it is not a true insulator at all. From the
diagonalization of the Hamiltonian of Eq. (\ref{HProxi}) we get the band
dispersion
\begin{equation}
\varepsilon_{\pm}= \frac {\Delta_{so} \tau _z s_z \pm \sqrt{\Delta_{so}  ^2 +
\frac { p^4}{ 4 m ^2}}} 2 \, \, .
\end{equation}
This band structure, shown in Fig. \ref{Bandas_proxi}, corresponds to an
indirect zero band gap semiconductor. The absence of SOC in one of the layers breaks the inversion symmetry,
making the bands non-degenerate  under independent
inversion of valley and spin.
When both valleys and/or both spins are considered, there is no gap
in the spectrum. However, for each spin and valley there is an
energy gap $\Delta _{so}$ at $\vec{p}=0$, and the system is thus an
insulator from an \emph{optical} point of view.  Moreover, if the
system is free from perturbations that mix the two valleys and the
two spins, it will also behave as an effective electronic insulator.
This has an important consequence on the existence of surface states
in the bilayer structure. In zigzag edge terminations for  which  it
is not necessary to admix valleys to satisfy the boundary
conditions\cite{Brey_2006a}, states coming from different valleys
can be treated independently and dispersive surface localized
channels appear in the spectrum, see Fig. \ref{edge_proxi}(a).
However, for armchair terminated ribbons where, in order to satisfy
the boundary conditions, the wavefunctions have to be a linear
combination of both valleys,  no localized surface states appear in
the band structure, see Fig. \ref{edge_proxi}(b). The absence of a
full energy gap and the lack of surface states in some boundary
terminations are clear indicators that a graphene bilayer with SOC
in only one layer is not a quantum spin Hall system.

\section{Conclusion}
In this work we reviewed the description of single layer graphene
with SOC as a topological insulator in terms of the first Chern
number, which arises naturally in the computation of the Hall
conductivity.  We showed then that the Chern number per spin in
bilayer graphene is two, twice that of the monolayer. This doubling
is reflected also in the number of topological surface states, as a
consequence of the bulk-surface correspondence rule. By numerically
computing the spectrum of finite size samples, we checked that
bilayer systems have twice as many edge states as a monolayer. This
was furthermore confirmed analytically in the case of surfaces with
armchair termination. The fact that the total Chern number per spin
is even means that the bilayer system is a weak topological
insulator, susceptible to gap opening if the system is subject to
spin-mixing perturbations (such as a weak Rashba SOC). We also
assessed the general stability of the bulk topological insulating
state of bilayer graphene with respect to Rashba SOC, a staggered
sublattice potential, interlayer bias, and trigonal warping. The
first three perturbations compete with the intrinsic SOC and, when
sufficiently large, spoil the inverted gap property that is crucial
to making the system a topological insulator. Finally, we examined a
bilayer graphene system in which only one layer has intrinsic SOC.
Although in this system individual valleys have  non-trivial Chern
numbers, the spectrum as a whole is not gapped, so that the system
is not a topological insulator.

{\sl Note added:} While this manuscript was in the final stage of
preparation, a manuscript by Cortijo, Grushim and Vozmediano
appeared\cite{Cortijo_2010} which studies the Chern Simons
coefficients in graphene and bilayer graphene using an effective
action formalism.

\section*{Acknowledgments}
Funding for the work described here was provided by MICINN-Spain via
grants FIS2009-08744 (EP and LB) and FIS2008-00124 (PSJ), and by the
NSF through Grant No. DMR-1005035 (HAF).


\begin{thebibliography}{33}
\bibitem[{not({\natexlab{a}})}]{note1}
\bibinfo{note}{For simplifying the discussion, here we consider systems for
  which the $z$-component of the spin commutes with the Hamiltonian.}
\bibitem[{not({\natexlab{b}})}]{nota_elsa}
\bibinfo{note}{If in the bilayer system, apart from the SO term, we also
  include a mass term coming form a bias voltage $V$ between layers, the
  eigenvalues would depend on $s_z$ and $\tau_z$ in the form $[\varepsilon
  ^{(\alpha)}]^2=v_ F ^2 p ^2 + \Delta _{so} ^2 +\frac{V^2}{4} \frac {\gamma_1
  ^2 } 2 \\ + (-1) ^{\alpha} \sqrt {\gamma_1^2(v_F^2p^2+\tau_z s_z V
  \Delta_{so})+V^2(v_F^2p^2+\Delta_{so}^2) + \frac {\gamma _1 ^4}{4} }$.}
\end{thebibliography}

\end{document}